# Excitonic emission in van-der-Waals nanotubes of transition metal dichalcogenides

*T.V. Shubina[1*], M. Remškar[2], V.Yu. Davydov[1], K.G. Belyaev[1], A.A. Toropov[1], and B. Gil[1,3]*

Corresponding Author: *shubina@beam.ioffe.ru

[1]Ioffe Institute, St. Petersburg, 194021, Russia

[2]Jožef Stefan Institute, Ljubljana, Slovenia

[3]L2C, UMR 5221 CNRS-Université de Montpellier, F-34095, France

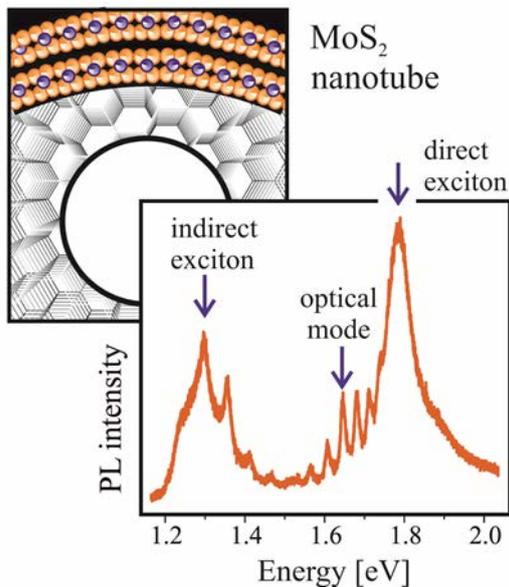

Nanotubes (NTs) of transition metal dichalcogenides (TMDs), such as $MoS_2$ and $WS_2$, were first synthesized more than a quarter of a century ago; nevertheless, many of their optical properties have so far remained basically unknown. This review presents the state of the art in the knowledge of the optical properties of TMD NTs. We first evaluate general properties of multilayered TMD crystals, and analyze available data on electronic band structure and optical properties of related NTs. Then, the technology for the formation and the structural characteristics of TMD NTs are represented, focusing on the structures synthesized by chemical transport reaction. The core of this work is the presentation of the ability of sinthesized TMD NTs to emit bright photoluminescence (PL), which has been discovered recently. By means of micro-PL spectroscopy of individual tubes we show that excitonic transitions relevant to both direct and indirect band gaps contribute to the emission spectra of the NTs despite having the dozens of monolayers in their walls. We highlight the performance of the tubes as efficient optical resonators, where confined optical modes strongly affect the emission. Finally, a brief conclusion is presented, along with an outlook of the future studies of this novel radiative member of the NTs family, which have unique potential for different nanophotonics applications.

# 1. Introduction

Allotropes of a 2D material, which are either a flat monoatomic layer or such a layer folded into a nanotube (NT), can possess essentially different optical properties. A well-known example of that are graphene and carbon NTs, representing various allotropes of carbon.[1] One could expect similar behavior for inorganic NTs made from transition metal dichalcogenides (TMDs), such as $MoS_2$ and $WS_2$, in which the monolayers are held together by weak van der Waals (vdW) bonds.[2-7] Interestingly, the carbon NTs were discovered as a new physical object in 1991.[8] The TMD NTs synthesized from the vapor phase were reported around the same time, in 1992.[2] However, their fates turned out to be completely different. After the initial elucidation of general physical properties of both, the interest to the carbon NTs was first stimulated by the discovery of luminescence from an individual tube,[9] and warmed up further by the generation of quantum light.[10] Naturally, that paves a road to their promising applications in nanophotonics and quantum technology.[11]

As for the synthesized TMD micro- and nanotubes (for short, we use further the notation "NT" for both when sizes are not important), they were proposed for different applications in mechanics,[12] in optoelectronics as detectors.[13] in electronics as transistors,[14] and also in tribology.[15] Similar to other inorganic NTs, they can be used in nanofluidics for non-destructive interaction with a living cell.[16] However, their application in nanophotonics as light emitters was never discussed because of the lack of data on their luminescent properties. It was not clear until recently whether they can emit light at all, not to mention what kind of radiation they could exhibit: either intense direct band gap exciton emission, as from a single TMD monolayer,[17] or weak emission of a momentum-forbidden indirect exciton, as from a bulk TMD crystal.[18] It looks unusual in relation with nanostructures made of the TMD 2D materials, which became so popular after the observation of the indirect to direct band gap crossover in the monolayer limit yet in 2010.[19]

For other vdW tubular structures, the situation was different. For instance, the emission was observed from HfS$_2$ NTs,[4] boron nitride NTs,[20] and also from the nanoscrolls fabricated by rolling up the MoS$_2$[21] and WS$_2$[22] monolayers. On the contrary, the studies of chemically synthesized MoS$_2$ and WS$_2$ NTs were focused on such characteristics of tubes ensembles as absorption[23] and optical limiting, i.e. nonlinear saturation of light transmittance at high incident intensities.[24] Theoretically, it was predicted that TMD NTs, independently on folding, should be semiconducting with a band gap smaller than in a bulk crystal.[25, 26] Also, it was noticed that multiwalled NTs must inherit the optical properties of the bulk crystal, i.e. show weak radiation characteristic of momentum-forbidden indirect band gap transitions. This prediction is in line with the common opinion on the low radiative ability of TMD layered crystals that is supported by experimental data.[18] Contrary to these anticipations, bright excitonic emission from the individual synthesized NTs has been recently discovered at the Ioffe Institute.[27, 28]

Here, we review the available data on technology, structural properties, and optical properties of TMD NTs. The layout of this paper is as follows. In section **2**, basic optical properties of layered vdW TMD crystals and related nanostructures are described. We believe that this collection of data on thin multilayer structures can be useful for understanding the optical properties of multiwalled NTs. In section **3**, we consider the electronic band structure and the previously known optical properties of synthesized TMD NTs as well as related curved nanostructures and artificial tubular structures. Section **4** presents the formation technology and the structural characteristics of TMD NTs, mostly for those that are synthesized by chemical transport reaction (CTR). We describe some of the experimental details of our micro-spectroscopy studies of individual NTs in section **5**. The recently discovered radiative properties of the single NTs are reviewed in section **6**, where we present micro-photoluminescence (mPL) data obtained with spatial and temporal resolutions. They indicate that the emission spectra of NTs contain, as a rule, two bands related to the radiative recombination of both direct and indirect band gap excitons. We underline that the direct exciton emission is bright enough despite the large thickness of the walls of the tubes containing several dozens of monolayers. In section **7**, we demonstrate that the vdW tubes can act as optical resonators and that the emission of both bands can be selectively enhanced by coupling with confined optical modes. The conclusions and outlook of the NTs studies are presented afterwards.

## 2. Multilayer $MoS_2$ and $WS_2$ crystals and nanostructures

The studies of TMD layered crystals and nanostructures have a history of more than half a century. The primary results can be found in the review of Wilson & Yoffe.[29] The TMD family comprises about 40 compounds of $MX_2$ type, where M is a transition metal and X represents a chalcogen. The slabs of three atomic layers X-M-X, with internal covalent bonds, are linked together by weak vdW forces. The TMDs crystallize into different polytypes, determined by the slab stacking and the coordination symmetry of the metal atoms. The hexagonal 2H polytype contains two slabs with the second one rotated with respect to the first by $\pi/3$ around the axis perpendicular to the stacking. In the particular 2Hb case, the layers in the unit cell are displaced so that the M atoms of one layer are directly above the X atoms of the other layer. The repeating cell of the rhombohedral 3R polytype composes the three slabs mutually translated in plane with a shift. In both, 2H and 3R, the metal atoms are surrounded by chalcogens forming a deformed trigonal prism coordination. The 1T polytype composes of one slab only but with the octahedrally coordinated metal atoms. The detailed description of the polytypes can be found in several reviews.[30,31] In this work, we focus on two layered semiconductors – $MoS_2$ and $WS_2$. Their natural crystals have 2H preferable stacking, while the crystals grown by transport reaction can have the 3R stacking as dominant.[29] Most of the published data on optical properties concern $MoS_2$ which is a stable and widely spread compound. Probably, this is also because the lowest dark exciton states in $WS_2$ can complicate the analysis of optical data.[32] Besides, $MoS_2$ has a particular significance for the modern 2D physics following the experimental observation of the direct band gap in its monolayers.[19]

Theoretically, the electronic band structure of bulk MoS$_2$ was analyzed using first-principles calculations. It was predicted that the band gap is indirect with a value of ~1 eV[33] in reasonable agreement with experimental data. The modification of the band structure from the bulk to the monolayer results from the strong quantum confinement in 2D materials.[34] The modern view on the band structures of MoS$_2$ mono- and multilayers with 2H stacking can be found, e.g., in Refs.[35,36]. In contrast to the monolayer case, where the optical transitions between valence band maximum and conduction band minimum take place around the K point at an energy of ~1.9 eV, in the bulk the momentum-indirect transitions Γ−K and Γ−Λ ( Λ is a midpoint between Γ and K) provide the onset of absorption edge at ~1.3 eV.

The change of stacking orders affects both electronic spectra and vibrational properties of layered crystals. The direct observation of valley-dependent out-of-plane spin polarization in 3R MoS$_2$ was done using spin- and angle-resolved photoemission spectroscopy. This result was ascribed to a non-centrosymmetric structure of 3R crystals, different from the centrosymmetric situation in 2H.[37] Raman studies demonstrated that interlayer interaction is weaker for the 3R phase, while intralayer vibrational modes are almost identical.[38] The first principles calculations showed that symmetry prohibits the splitting of the degenerate states in the bands of bilayer, while this effect appears in the triple layers of MoS$_2$ with 3R stacking.[39] In contrast to the 2H stacking, the Γ−K rather than Γ−Λ transitions have the lowest energy in such 3R structures. The triple layers can be considered as the onset of "bulk"; thus such characteristics seem to be typical for the bulk 3R structures. (We pay attention to this fact because the 3R folding dominates in the synthesized NTs.)

Most of previous optical studies of the multilayer TMD crystals were done using rather thin samples with a thickness in the 50-500 nm range. These samples were formed by repeated cleaving on an adhesive tape, i.e. by technique which is similar to the exfoliating method currently used. Strong A and B direct band-gap excitonic resonances (two – because the valence band is split by the spin-orbit coupling) were observed in absorption spectra of such MoS$_2$ samples around 1.9 eV.[29,40] Similar features in WS$_2$ are somewhat higher, at ~2 eV. The absorption coefficient of MoS$_2$ in the region of direct exciton resonances is in the (5.2-5.7)*10$^5$ cm$^{-1}$ range.[40] The absorption spectra exhibit a shift of about 100 meV when the temperature rises from 77 K to 300 K. Samples with 3R stacking show the excitonic peaks at slightly lower energies.[29] The bulk crystals can reflect 20% of impinging light. The optical properties of them turn out to be strongly anisotropic. Data on the ordinary and extraordinary refractive indices in MoS$_2$ can be found, e.g., in Ref.[40]. We highlight that no marked optical peculiarities were found in absorption and reflection in the spectral range of weak indirect excitons.

To the best of our knowledge, the first studies of the radiative properties of 2H MoS$_2$ bulk crystals, both synthesized and natural, in the vicinity of indirect excitons were reported by Kulyuk et al. much later, in 2003.[18] The measured PL spectra contained narrow excitonic lines and their phonon replicas in the 1.15-1.3 eV range, as well as a broad band centered at 1.0 eV, most likely related to defects. The temperature evolution of the PL spectra revealed the fast quench of the excitonic lines: they almost disappeared at T>50 K. The total PL intensity decreased by two orders of magnitude from 2 K to room temperature. Kulyuk et al. ascribed the 1.15-1.3 eV emission to the halogen transport agent intercalated during the growth process, because such emission was absent in the studied natural crystals. Note that the spectral range of direct excitons has not been studied, while this might be of interest, since, in addition to doping, intercalation is an effective way of separating monolayers from each other,[41] and direct exciton luminescence at higher energies could appear. Thus the influence of impurities on optical properties needs additional studies.

The discovery of nanocrystal quantum dots in 1980$^{th}$ [42] has stimulated the investigation of quantum-size effects in MoS$_2$ nanocrystals. The synthesized MoS$_2$ nanoclusters of 2-15 nm in size (3-25 monolayers), suspended in a nonpolar fluid, exhibit distinct spectral features in absorption, which shift towards higher energy with decreasing the cluster size due to the quantum confinement.[43] The splitting between A and B excitonic peaks increases as well. The lowest energy PL peak in such nanoclusters was registered near 2 eV, which is very close to the energy of direct exciton transitions.[44] Time-resolved PL (TRPL) measurements of the $MoS_2$ nanoclusters were performed using a time-correlated single photon counting technique.[45] It was shown that the radiative recombination is controlled by trap-to-trap relaxation. The emission decay is relatively slow (~200 ps) at 20 K and becomes faster (<40 ps) at room temperature that may evidence an effective carrier transport towards non-radiative centers. Note that the exciton radiative lifetime in a TMD monolayer is very fast, of the order of few picoseconds.[46]

About ten years later, $MoS_2$ triangular nanoplatelets of several nanometers in lateral dimensions were fabricated. These particles keep the planar geometry with their size growing; the dangling bonds at their boundaries are stabilized by excess S atoms.[47] The study by scanning tunneling microscopy (STM) exhibited structural transitions taking place at certain cluster sizes.[48] Theoretical studies of nanoplatelets by ab initio calculations have demonstrated the possibility of band gap variation in a wide spectral range from indirect to direct excitonic transitions in dependence on the number of sheets and their interspacing.[49] Such small particles are assumed to be promising for tribology.[50]

In the common opinion, the PL related to direct excitons should be completely quenched in the TMD bulk crystals and multilayered nanostructures. Indeed, there are experimental data showing a dramatic decrease down to full disappearance of luminescence efficiency in the bulk (several-layers samples) as compared with a monolayer.[17] It contradicts other studies of flakes comprising a few layers, where both lines of direct and indirect excitonic emission coexist in PL spectra. For instance, to confirm the direct-indirect band gap crossover in $MoS_2$ Mak *et al*. have measured both direct and indirect exciton lines in samples which contain up to 6 monolayers, i.e. almost "bulk".[19] It is interesting to emphasize that the direct exciton line still dominated the PL spectrum in flakes with several monolayers and that its intensity was only one order of magnitude weaker than it is in a direct-gap monolayer. Similar data have been presented by Dhakal *et al*.[51]

There are several factors that can promote the direct exciton emission in a multilayer TMD structure. Among them is a temperature rise, which can push the system toward the 2D limit by decoupling the neighboring layers with interlayer thermal expansion.[52] In this case, PL in the multilayer flake exhibits a decrease in intensity by three times only from 100 K to 400 K. Such robustness might be rather typical for a monolayer. Impurities intercalation can also provide the layer separationl.[41] Elastic strain affects strongly optical properties as well (see for review Refs.[53, 54]). The local strain arises due to either layer extension or curvature; the latter is characteristic for nanotubes. Reduction of the direct band gap and formation of the localization sites was observed on the top of intentionally formed riffles ("semi-tubes") in the monolayers.[55] Most of the studies of strain effects were done using $MoS_2$ monolayers.[56-60] It was shown that the tensile and compressive strains act in the opposite ways, enhancing either direct or indirect transitions, respectively.[57] PL spectroscopy demonstrated that the band gap can be tuned in a continuous and reversible way by 500 meV under biaxial strain.[58] Importantly, it was shown that 2D materials sustain very large deformations without damage. That is why the strain is currently considered as an effective way for the modification of the band structure [59] and the enhancement of nonlinear optical response.[60] In both cases, the experimental confirmation was obtained by curving the monolayer samples positioned on a flexible polymer substrate. These results are apparently useful for the consideration of nanotubes properties.

Recently, there appeared the trend to consider exciton emission dynamics by taking into account momentum forbidden exciton states. It concerns even the TMD monolayers where the direct gap exciton emission dominates.[61] Besides, it was recognized that the dynamics is dependent on whether dark or bright excitonic state sits at the lowest energy in a particular material.[32] It is worth mentioning that the relaxation process of photoexcited carriers in 2D materials strongly depends on the created carrier population and, hence, on the energy of excitation. The closer is the excitation energy to the direct exciton resonance, the more pronounced is its performance.[51,57,62] Thus, we assume that in a multilayer system there is always an optimal balance between several competing channels of exciton creation, relaxation, and recombination related to interband transitions between various valleys.

**3. Basic optical properties of TMD nanotubes and related curved nanostructures**

Electronic band structures of both $MoS_2$[25] and $WS_2$[26] NTs have been studied using the density-functional-based tight-binding method. The primary works considered the nanotubes as a single-wall cylinder constructed by conformal mapping of a triple sheet (X-M-X) with different folding: armchair (n=m), zigzag (n≠0, m=0), and chiral (n≠m). It was found that $MoS_2$ NTs are mechanically stable. Independently of the folding, all these NTs are semiconducting, in contrast to the carbon NTs where the folding type determines either metallic or semiconducting character. The value of a NT band gap rises with increasing the diameter, D, approaching the monolayer limit, because the strain tends to be smaller and smaller with that, following roughly the $1/D^2$ dependence. Experimentally, a confirmation of the band gap dependence on the tube diameter was obtained by scanning tunneling microscopy (STM) measurements of I-R curves.[63] Using Raman excitation was another tool to determine the band gap variation as a function of the NT diameter.[64] The particular band-gap character in the single-wall NTs was predicted to depend on the folding type. Namely, the armchair tubes should have small indirect and a moderate direct band gap values, whereas the band gap is proposed to be definitely direct in the zigzag tubes.[65] It is interesting to note that this is the first prediction of the direct gap in the monolayer limit, although done for a single-wall $MoS_2$ NT. Moreover, it was underlined that the NTs possessing the direct-gap transitions are very promising for light emitting devices.

Symmetry-based density functional theory (DFT) calculations of the electronic spectra of single-wall $MoS_2$ and $WS_2$ NTs [65, 30] enriched the results of Seifert *et al*. as follows: the direct band gap in a zigzag tube has the same diameter dependence as the indirect band gap in an armchair tube. With increasing the diameter, the indirect and direct gap values in the $MoS_2$ tube tend to 1.15 and 1.3 eV, respectively. The highest limit for the direct gap of $WS_2$ NTs is about 1.65 eV. Note that the estimated gap values are markedly lower than experimental ones. For instance, low-temperature absorption spectra have peaks around 2.1 eV in both WS$_2$ bulk and a thin coating layer comprising arrays of NTs of 15-20 nm in diameter.[23]

As expected, the curvature-induced strain in the TMD NTs is dependent on the tube diameter, namely, the smaller the diameter the higher the strain. The curvature strain shifts the calculated spectra of optical absorption toward lower energies in unrelaxed MoS$_2$ NTs, whereas the strain relaxation leads to the shift of absorption peaks towards higher energies.[66] The concept of the strain influence on optical properties was later developed in a set of papers.[67-70] The DFT calculations showed that the properties similar to those in the monolayer and bulk are realized, respectively, in a large diameter single-wall NT and a multiwalled NT. It was predicted that the strain can transform the direct band gap into an indirect one. Such transition can take place for condition of the tube elongation by as much as of 16%.[66]

The Raman signals of the in-plane and out-of-plane lattice vibrations should also significantly depend on the strain value. Thus, the Raman spectroscopy is a unique method to determine the strain-dependent characteristics of individual NTs. A review of theoretically evaluated phonon dispersions in the TMD tubes can be found in Ref.[30, 31] The multiwalled WS$_2$ NTs were investigated by confocal micro-Raman spectroscopy. It was found that the inter-layer vibration mode, $A_{1g}$, is sensitive to both the curvature-induced strain and diameter, while the in-plane vibration mode, $E^1_{2g}$, is not affected by them. The competition between the vdW force stiffening and the curvature strain softening provides the shift of the $A_{1g}$ mode frequency by 2.5 cm$^{-1}$ in the multi-layered nanotubes with small diameters.[69] The micro-Raman technique was also exploited to compare the structural and vibrational properties of nanoplatelets and nanotubes.[70]

Optical absorption and light reflection/scattering in the NTs turn out to be highly anisotropic. In all previous studies, the prevailing signal was polarized along the nanotube axis. This was explained within the formalism of antenna theory, assuming that the depolarization effect in a nanotube strongly suppresses the light polarized perpendicular to the nanotube axis.

Such effect was investigated by precise Raman measurements of an individual $WS_2$ nanotube suspended on a cantilever.[71] The calculated polarized spectra were well consistent with this concept.[65] We believe that this depolarization effect can concur sometimes with the action of optical modes which are polarized along the tube axis as well.[28] One more interesting feature of NTs is the nonlinear saturation effect (optical limiting) which is stronger than those in the carbon NTs. In the $MoS_2$ NTs in aqueous suspensions, this effect results in decreasing the transmittance by an order of magnitude with the increasing power of 532-nm and 1064-nm nanosecond laser pulses.[24]

A set of nanostructures with a curved semicircular surface should be attributed to the family of the tube close relatives. The most interesting example is represented by the nanostructures formed by deposition of a $WSe_2$ monolayer onto patterned silica substrates with the arrays of pillars of 150 nm in diameter and a 60-190 nm height.[72] The deposited monolayer is curved on the pillars, forming quantum-dot like localization sites at pillars tops due to local strain. These sites turned out to be the effective emitters of single photons, as it was confirmed by the correlation measurements of narrow excitonic PL lines. These narrow lines were recorded in the spectral range of ~50 meV; thus, a large-scale set of relatively uniform single-photon emitters can be formed using such approach. Another interesting example is nanoscrolls produced by rolling up a $MoS_2$ monolayer within a drop of ethanol.[21] The direct exciton emission from such nanoscrolls was registered with excitation by a 532-nm laser line. The energy of emission is slightly lower than in a monolayer in agreement with the theoretical prediction of the strain-induced shift.[65] In this work, the spectral range of indirect exciton transitions was not investigated; thus there is still uncertainty whether the indirect gap emission exists in such nanoscrolls. The optoelectronic performance of $WS_2$ nanoscrolls was investigated by rolling up 2D sheets and fabricating p-n junctions out of the formed tubular structures.[22] Light-emitting diode and photovoltaic effects were observed in such tubular structures, named conditionally as nanotubes.

Remarkably, it was reported that $WS_2$ NTs can sustain both excitonic features and a cavity mode in the visible – near infrared range.[73] This result was obtained using an ensemble which represents the aqueous dispersion of 1-10 µm long synthesized NTs of 40 – 150 nm in diameter with 20–40 shells in their walls. The comparison of spectra of the absolute absorption and standard extinction (the later includes both absorption and scattering) have revealed an extra dip arising, presumably, due to the interaction between the excitons and the cavity mode. This observation was supported by performed finite-difference time-domain (FDTD) simulations. The absence of luminescence from the studied NTs was ascribed to the indirect band gap in $WS_2$. We should note that the direct observation of optical modes modulating the excitonic PL spectrum has been recently reported.[27] This phenomenon will be described in section 7, while in section 6 we present the evidences of PL in multilayered synthesized $WS_2$ NTs.

**4. Synthesis of TMD nanotubes**

The first reports on the existence of $WS_2$ and $MoS_2$ crystals with cylindrical and spherical shapes were published, respectively, by Tenne *et al*. in 1992 [2] and Margulis *et al*. in 1993.[74] The samples were prepared by sulphurization of the respective metal oxides acted as self-sacrificed precursor crystals, which determined the aspect ratio of the final products. Microtubes and nanotubes of $MoS_2$ (Figure 1a) and $WS_2$ (Figure 1b) synthesized by chemical transport reaction (CTR) were firstly reported in 1996 [3] and 1998.[75] The synthesis of TMD nanotubes by different methods is discussed in a set of comprehensive reviews and books;[2, 4-7, 76, 77] here we focus on the tube formation by the well-developed CTR.

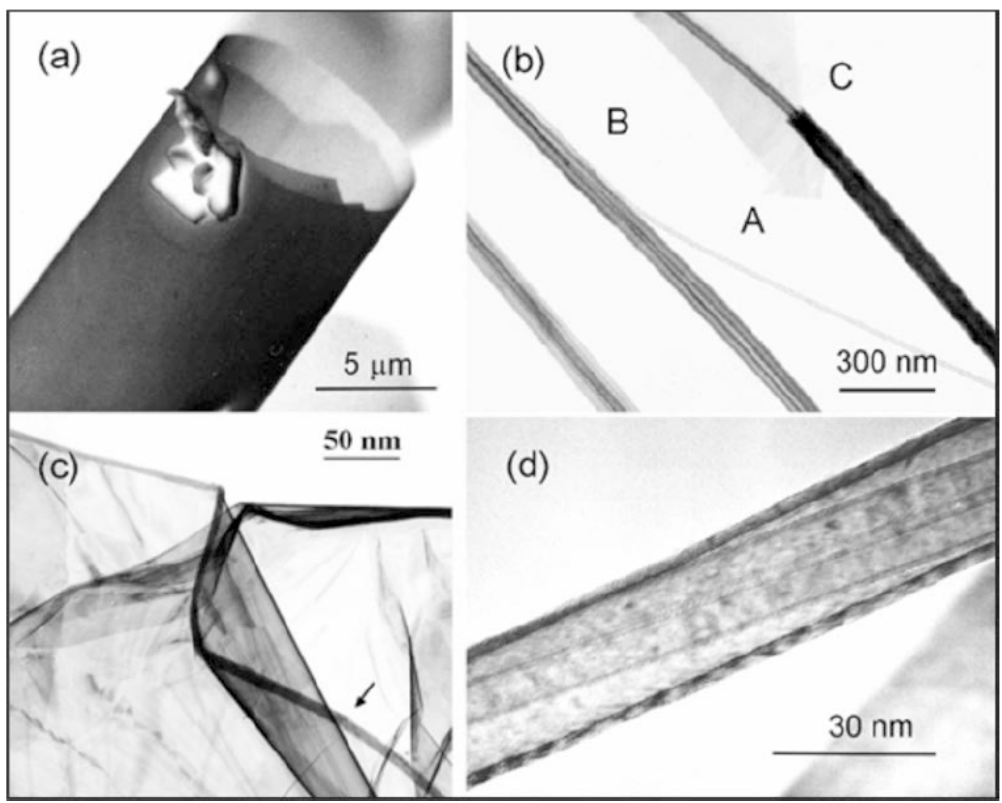

**Figure 1.** TEM images of different TMD tubes. a) A MoS$_2$ microtube, 11 µm in diameter and 90 nm in wall thickness; b) WS$_2$ nanotubes grown as single nanotubes (A), self-assembled into ropes (B) or co-axial nanotubes (C); c) nucleation of a nanotube at the edge of a thin 2D WS$_2$ flake; d) two WS$_2$ nanotubes rolled up around a central nanotube following its chirality.

The employed CTR growth mechanism was explained through instability of very thin flakes of layered materials against folding, which triggers formation of rolls from thin flakes or growth of self-terminated tubular structures inside microfolds or curved crystal edges (Figure 1c). The length of the rolls is in a range of a few tens of diameters and they grow only in radial direction in an orientation-disordered way. In contrast, the regular stacking between the molecular layers in the tubes enables their growth in longitudinal direction to a length, which extends with duration of a growth process and depends on parameters of the local environment, e.g. on size and shape of available space, diffusion paths of transported molecules, and vicinity of other tubes or flakes. The growth takes place directly from vapor phase. The tubes grow at a very slow rate and only at their very ends, where the unsaturated bonds represent the nucleation sites with minimal free energy. The molecules adsorbed on the surface of a tube with free vdW bonds are either transported along the tube towards its ends or they are desorbed back to the gas environment. Very homogeneous diameter and wall thickness indicate that nucleation of a next layer on the surface is not preferable and simultaneously evidence an absence of surface defects where such a secondary nucleation would start. How these adsorbed molecules diffuse along the surface of tubes, namely, whether they follow the chirality of the surface atoms, or are hopping to the very end of a tube in straightforward way, is an open question at the time being.

The $MoS_2$ and $WS_2$ nanotubes are synthesized in closed silica ampoules using iodine as a transport agent. During the process of CTR, iodine reacts with the transition metals at high temperature, forming a volatile product, which decomposes at lower temperature, where transition metal reacts again with sulphur, forming solid transition metal disulphide.[78] The CTR runs from the high temperature zone (1125 K) to the low temperature one (1060 K) with a temperature gradient of 5.6 K/cm. After three weeks of growth, the silica ampoules are slowly cooled down to room temperature with a controlled cooling rate of 15°/hour.[3] Approximately a few percent of the starting material is transported in a shape of nanotubes, while the rest of the transported material grows as strongly undulated thin plate-like crystals. The lasting nearly equilibrium growth conditions enable the synthesis of nanotubes of very different sizes, from 10 nm to a few µm in diameter and from a few µm to several hundred microns in length. The tubes always grow together with large area thin flakes, which are typically rotationally disordered about the plane normal. The advantage of this kind of synthesis is an extremely low density of structural defects in the crystal structure of the tubes. It is worthy to mention that CTR mimics crystal growth in nature, where minerals grow in limited space and with limited available resources. Partial pressures of the different elements change during the growth, which affect equilibrium conditions. The nanotubes grow up to the end of duration of the growth run and cover the flakes of undulated crystals like spider nets.

While $MoS_2$ micro and nanotubes usually grow as single tubes, the $WS_2$ tubes tend to form ropes with the spontaneous self-assembly of the nanotubes during the growth.[79] Fluctuations of charges and consequent electric attractive forces were proposed as an origin of this attraction. Nucleation of the tubes at edges of thin flakes, inside microfolds, and at weakly bond surface steps leads to a nucleation of several nanotubes grown side by side.[80] A possible nucleation site is also a surface step at the inner or outer surface of a hollow tube, which enables for a growth of more complex tubular structures, like co-axial tubes (Figure 1b), or mutually rolled up nanotubes, where each of the nanotubes follows the lattice structure of the adjacent nanotube [75] or of the wider central tube (Figure 1d). In the tubular shape of $MoS_2$ and $WS_2$, the interlayer distance is extended in average by up to 3% with respect to the $2H_b$ bulk value.[3] This extension differs with diameter and number of the layers composing a wall of tube. The extension is the largest near the outer surface, while the separation among the layers shrinks in direction toward the center of a tube to the bulk value or even to slightly lower value. This leads to mechanical instability in the central layers, which are therefore exposed to breakage.[5] The reason for this is a strain incorporated in the tubes walls, which is caused by the interlayer stacking and curvature. Although the stacking polytypes were defined for flat geometry, one can speak about the stacking order of the layers in tubular shape too, at least in parts of the tubes. The cylindrical geometry cannot be formed from flat layers without consideration of strain in extended and contracted triple S-Mo-S or S-W-S layers, and/or creation of defects, which relax this strain, when it exceeds a critical value.

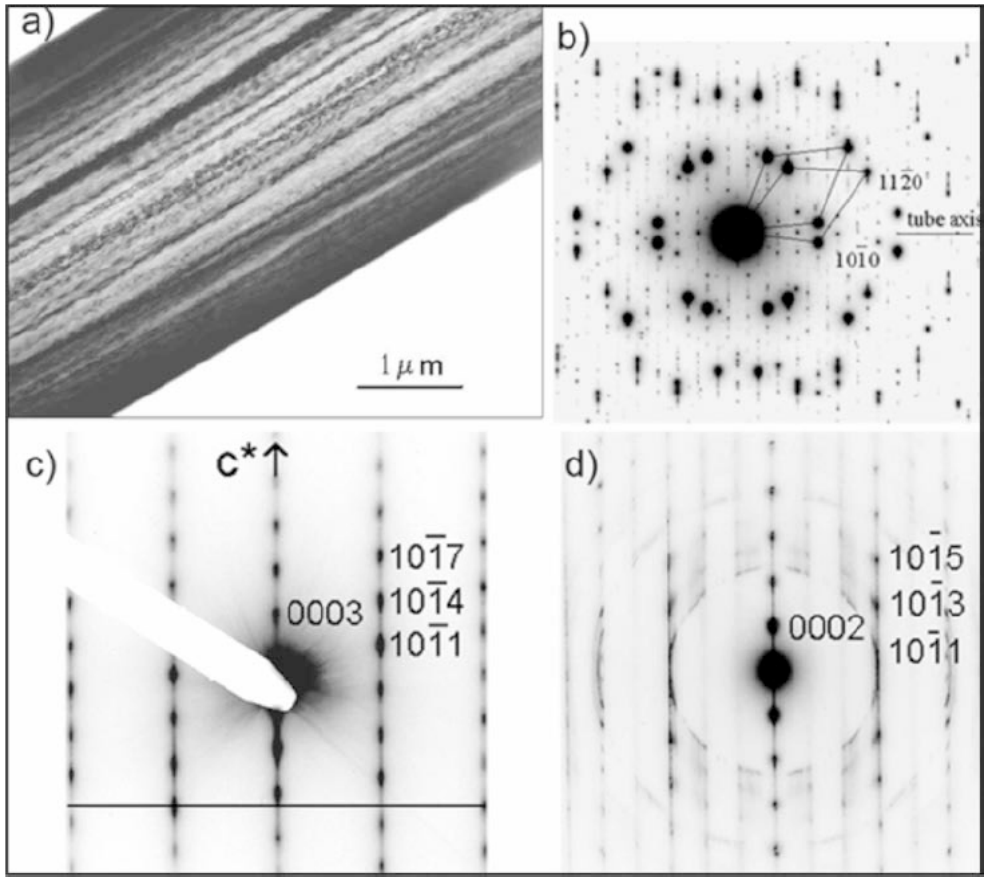

**Figure 2.** a) MoS$_2$ microtubes with bending contours along the MoS$_2$ microtube axis revealing strips of semi-long-range order, satisfying Bragg's condition for electron scattering. Electron diffraction patterns taken: (b) at the central part of the tube as a superposition of electron scattered by upper and lower walls revealing the chiral growth mode; (c) at the tube edge with the rhombohedral 3R stacking sequence in accordance with the rule (-h+k+l=3n); (d) on a nanotube, 190 nm in diameter, crystallized in the 2H$_b$ stacking sequence satisfying the rule l=±(2n+1).

The tubes grow in chiral growth mode with angle of chirality between projection of the tube axis on the [0001] basal plane and the $<10\bar{1}0>$ lattice direction (Figure 2a,b). Microtubes with diameters around a few µm grow in orthorhombic (3R) stacking (Figure 2c), while more narrow nanotubes are crystallized in hexagonal (2H$_b$) stacking (Figure 2d).[80] This difference influences the optical properties of the nanotubes due to in-built lattice strain in the microtubes, which can cause up-shift of Raman peaks with respect to the bulk.[81] The lattice parameters in bulk 3R polytypes are for approx. 1% (MoS$_2$) and 2% (WS$_2$) larger in basal plane, and slightly (0.2%) shorter perpendicularly to the layers.[29] In microtubes, the strain is distributed over a larger number of primitive unit cells composing the circumference of a tube. It stabilizes a high-pressure 3R stacking, which in bulk is stable at pressures above 40 kbar.[79] The strain intensity is increasing toward the central part of a tube that stabilizes the coexistence of several diameters over the whole length of a tube. As-grown tubes are open ended, but not all of ideal cylindrical shape. If a tube meets an obstacle during the longitudinal growth, it deforms to elliptical cross section or collapses to the shape of a ribbon, and then continues its growth in this new shape. Because of the chiral lattice structure, the strain at strongly bend molecular layers at longitudinal edges of the elliptical tubes or ribbons causes the spiral twisting of these forms.[82] Besides, the domains of various stacking can be observed in nanotubes grown by different methods. The influence of domains on structural properties was reported by L. Houben et al.[83] The effect of domains on optical properties requires additional research.

## 5. Experimental details and samples characterization

To investigate the optical properties of individual MoS$_2$ and WS$_2$ NTs we used two setups for micro-spectroscopy measurements, which have different functionality. Both setups exploited an optical confocal scheme. Each setup is equipped with a silicon liquid nitrogen cooled CCD for the mPL measurements in the visible – near infrared range (up to 900 nm). Micro-cryostats allowed the measurements at different temperatures (down to 8 K). Large working distance lenses (Mitutoyo 100×NIR, NA = 0.50) and pinholes ensured the light collection from a ∼2-μm spot on a sample that enables to measure mPL from a single tube.

One of these setups is equipped with piezo-drivers Attocube XYZ placed inside the helium cryostat that allowed us to set and hold the position of selected nanotubes during quite a long time. In particular, it is possible to measure mPL spectra focusing on different points along a tube to confirm its homogeneity. Besides the CCD, this setup also has a silicon avalanche photodiode which operates in a time-correlated single-photon counting mode that permit us to measure mPL with a time resolution of about 45 ps. The TRPL measurements of tubes and flakes were done at low temperatures (10 K) using for excitation a 404-nm (3.07 eV) line of a picosecond pulsed semiconductor laser. Power of the laser was 1 mW in average as measured in front of a cryostat window. The repetition frequency of the pulses was 100 MHz. To detect PL decay curves at selected wavelengths the collected emission was dispersed by a grating monochromator.

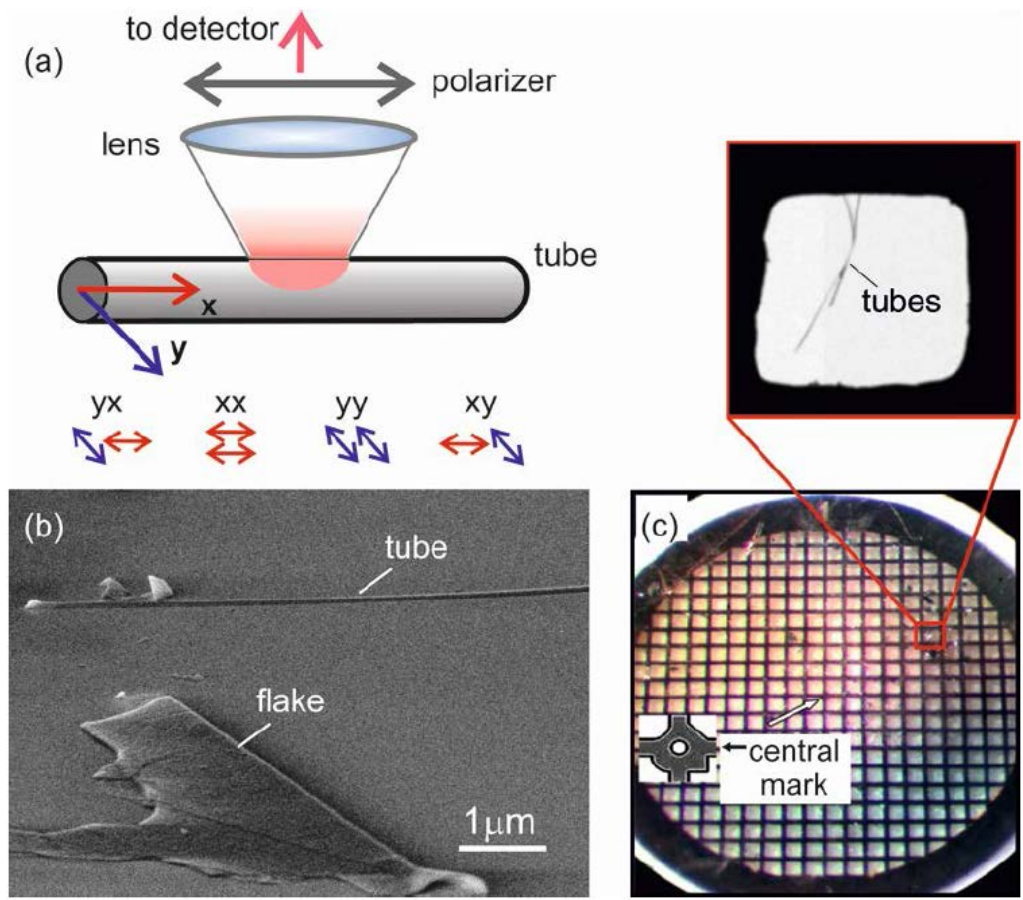

**Figure 3.** (a) Schematic of the excitation and detection of mPL from a single nanotube. Red and blue arrows in bottom indicate possible polarization configurations with respect to the tube axes. The notation used like "*yx*" reads as follows: excitation along the "y" axis, detection along the "x" axis. (b) SEM image of a nanotube and a flake on a silica substrate. (c) Square mesh TEM grid of ~2 mm in diameter with suspended tubes, ribbons, and flakes used in the experiments. The asymmetric central mark is useful to find a cell. The inset shows one cell with a higher magnification.

The other setup was primarily intended for micro-Raman (mR) spectroscopy. The Raman spectra were always measured to characterize quality of the studied samples. Besides, the intensity of Raman scattering was registered during the scans of PL bands that was used afterwards for normalization of the PL intensity as described in Ref.[17]. In this setup, the linear-polarization sensitive measurements of mPL were done in four possible configurations (see Figure 3a). This setup comprised a Horiba Jobin-Yvon T64000 spectrometer equipped with a confocal microscope and a silicon CCD. A temperature controlled microscope stage Linkam THMS600 allowed realization of temperature-dependent measurements. In addition to the CCD, the setup possessed an InGaAs-based linear diode array (sensitivity range 0.8–1.7 μm). Combination of two detectors allowed us to determine the true shape of indirect exciton emission, which is located spectrally just at the sensitivity boundaries of both. Besides, it enables to control the defect-related emission which might appear at ~1 eV.[18] In these continuous wave (CW) measurements, we used a 600 gr/mm grating and a 532 nm (2.33 eV) line of a Nd:YAG laser for excitation.

We have investigated both tubes and flakes grown in the same silica ampoule during the same growth run of CTR, as described in section 4. For reference, the multilayered flakes exfoliated from a commercially available high-quality $MoS_2$ bulk crystal ("HQ-graphene" production) were studied as well. Note that no appreciable difference in optical properties was observed between the synthesized and exfoliated flakes. In addition to the samples on a $SiO_2$/Si substrate (Figure 1b), we have studied the free-standing tubes and flakes that were suspended in a square mesh grid commonly used for TEM investigations (Figure 1c). We have previously exploited such approach to establish the link between mPL and structural defects in a quantum well structure.[84] We have produced samples with the suspended tubes using conventional tweezers without "tape" technology. The tube surface remains clear after manipulations, without any contaminations. Importantly, the use of free-standing samples excluded the possible effect of a substrate.[85-87] Besides, it permitted us to perform the TEM characterization of selected tubes to obtain information on their structural properties. It is also worth mentioning that there was no strong difference between general dependencies of PL in free-standing samples and samples placed on a Si substrate. However, we observed the strong difference between the tubes and flakes.

Before the mPL studies, the samples were characterized by micro-Raman spectroscopy. The typical room-temperature Raman spectra of NTs and multilayer "bulk" flakes are presented in Figure 4. The polarization dependency of the Raman active modes, with almost full suppression of $A_{1g}$ in the $yx$-polarization, evidences the good structural quality of CTR-grown NTs. Additional studies of exfoliated flakes of atomic thicknesses gave us the energy-thickness dependence of the Raman modes, which is similar to that described in literature.[88] The measured shifts of the $E^1_{2g}$ and the $A_{1g}$ modes with respect to their positions in a monolayer, indicated in Figure 4b, correspond to the "bulk" case, when the number of monolayers is definitely more than ten. In the low-frequency range, the shear mode at 32.5 cm$^{-1}$ is significantly suppressed in the tubes, as can be anticipated.[89] Thus, we confirm that all our tubes under study are multiwalled. The increase of the $A_{1g}/E^1_{2g}$ intensity ratio, along with the appearance of some contribution of a breathing mode at the low-frequency side of shear peak, is suggestive of 3R folding.[38] These findings are well consistent with previous TEM studies revealing the dominance of the 3R polytype in the relatively large multiwalled tubes grown by CTR (see section 4).

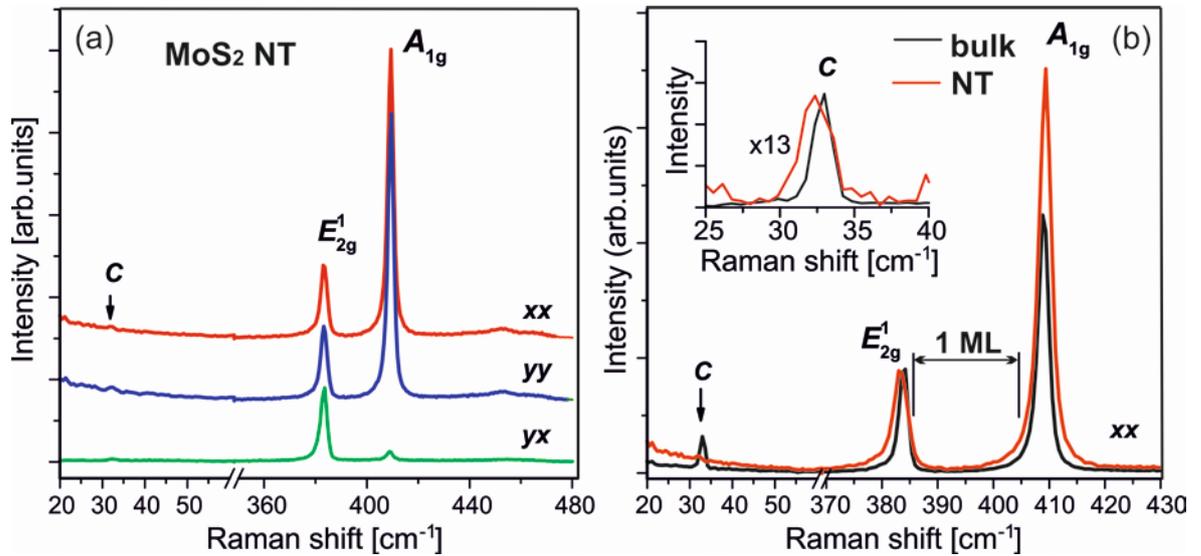

**Figure 4:** Raman spectra of a MoS$_2$ tube of 2 µm in diameter measured at room temperature using a 532-nm laser line excitation: (a) spectra taken in different polarization configurations marked in the plot; (b) comparison of Raman spectra measured in the tube and in a multilayered "bulk" flake. The position of the modes in a MoS$_2$ monolayer is marked. The inset shows the range of low-frequency modes.

## 6. Luminescence emerging from TMD nanotubes

We studied micro- and nanotubes of different diameters ranged from 400 nm to 2 µm and found that they are radiative. Characteristic low-temperature spectra of MoS$_2$ and WS$_2$ NTs recorded in the spectral range of direct excitons are displayed in Figure 5. Notable fact is the observation of bright direct exciton emission from the multilayered structures, which possess an indirect band gap. Moreover, the integral PL intensity in the MoS$_2$ NTs appears to be an order of magnitude higher than in the flakes fabricated during the same growth run. This estimation is done taking into account the ratio of excited surface areas, S, i.e. $S_{flake}/S_{tube}$~5-10. The intensity of emission from the WS$_2$ tubes is markedly lower, probably because of the non-ideality of the WS$_2$ NTs, discussed in section 3. These WS$_2$ NTs are likely relaxed because the strain-induced shift of transition energies was not observed. However, they resemble the MoS$_2$ tubes in many aspects such as the temperature behavior of PL and the performance of resonant optical modes (see section 7). This permits us to present here the results obtained using more stable and brighter radiating MoS$_2$ NTs, which can be considered as a typical representative of the TMD tube family.

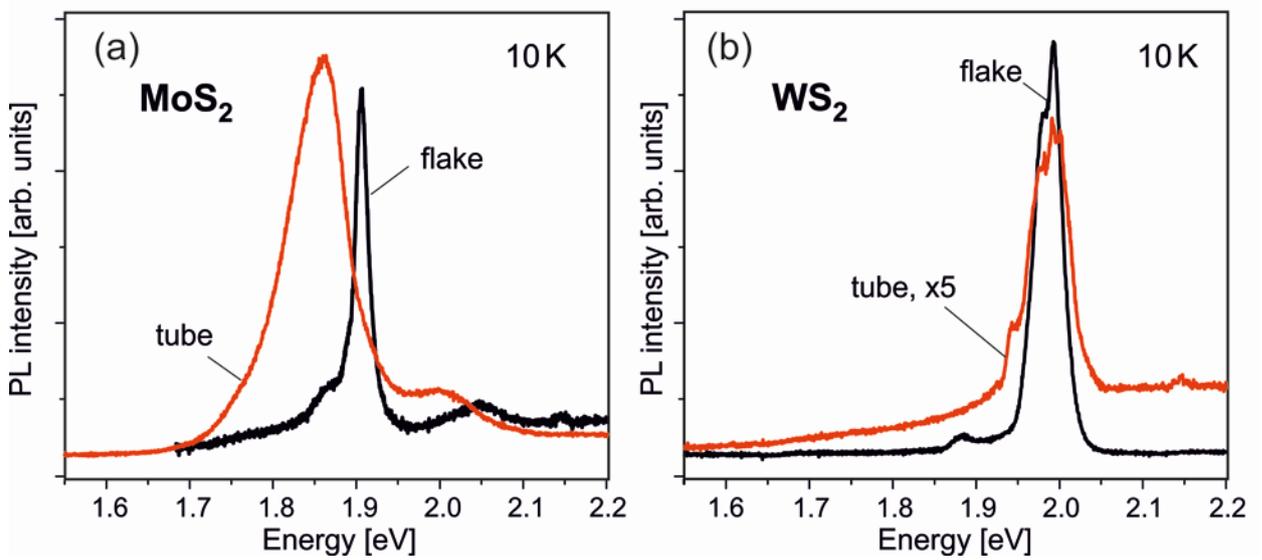

**Figure 5.** Spectra of mPL measured in the spectral range of direct exciton transitions at 10 K using 404-nm laser line excitation from: (a) MoS$_2$ and (b) WS$_2$ nanotubes with the diameters of ~400 nm. The non-ideality of WS$_2$ tube results in the lower PL intensity, while in the perfect MoS$_2$ tube the integral intensity can significantly exceed that in the flake.

At low temperature, the spectra of the MoS$_2$ NTs display peaks at 1.86 eV and 2.00 eV, related to A and B excitons, respectively. Their energies are close to those in planar atomic layers.[18] The red shift of the peaks by ~50 meV with respect to the emission from the flakes can result from the combined actions of unrelaxed strain in the tubular structures[65] and 3R folding type (the excitonic peaks are red-shifted in 3R MoS$_2$).[29] In contrast to the previous measurements of PL from bulk crystals, which quenches completely at temperatures above 50 K,[18] the emission from the CTR NTs exists up to room temperature. The common feature of MoS$_2$ tubes and flakes is rather weak emission at low temperature of a phonon-assisted indirect exciton, while the direct exciton related emission is very pronounced. Among obvious reasons for that, we should mention the low population of the phonon bath at helium temperatures and the possible decoupling of neighboring layers.[52] With increased temperature, one can observe two bands of direct and indirect excitons in the PL spectra of both tubes and flakes (Figure 6).

However, their dependencies with the temperature rise are different. The intensity of the direct-exciton PL drops with rising the temperature in both cases as it takes place in conventional semiconductors, whereas the PL intensity related to the indirect-exciton transitions increases. Moreover, the increase of the indirect-exciton PL in NTs is markedly weak as compared to the planar flakes. The unknown fine structure of the exciton states in NTs can be the cause of such a difference, which requires additional research.

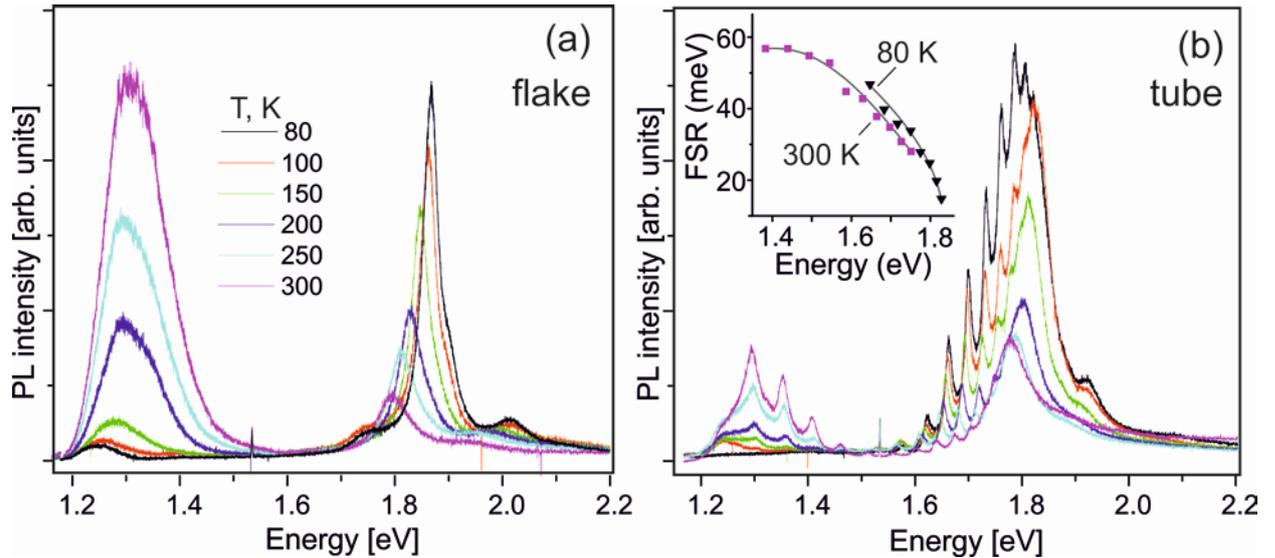

**Figure 6.** Temperature variation of mPL spectra in MoS$_2$ structures: (a) flake and (b) tube of 2 µm in diameter. The sharp peaks of WGMs modulate the spectra. The inset in (b) shows the energy dependence of free spectral range (FSR) between adjacent peaks, which follows the refractive index variation with temperature.

The characteristics of PL decay in the NTs resemble those of the radiative process in monolayers[90] rather than in bulk. The basic decay of the direct-exciton PL, measured at low temperature (Figure 7), turns out to be fast in both tubes and flakes. Its characteristic decay time is less than the temporal resolution of our system (<45 ps). A slowly decaying PL component, related presumably to defect states, decays almost similar in both tubes and flakes with decay time constant of 250 ps and 280 ps, respectively. We notice that the fast PL component in the NT decreases more rapidly and falls to the smaller value than in the flake, which likely evidences more effective recombination process of the direct band gap excitonic transitions in the NTs.

This assumption is consistent with the mPL spectra measured in a wide spectral range, which demonstrate relative suppression of indirect exciton emission in the NTs (Figure 6).

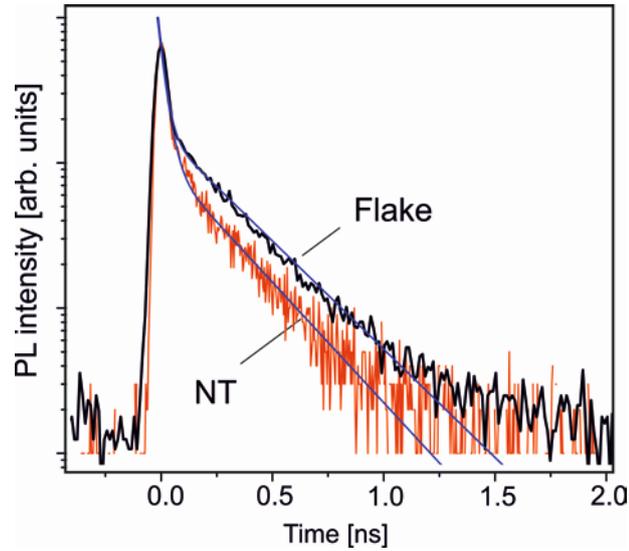

**Figure 7.** PL decay curves measured at 10 K at the peak energy of direct-exciton emission band in a MoS$_2$ nanotube of 500 nm in diameter (red line) and in a flake (black line). The solid lines show their two-exponential curve fittings.

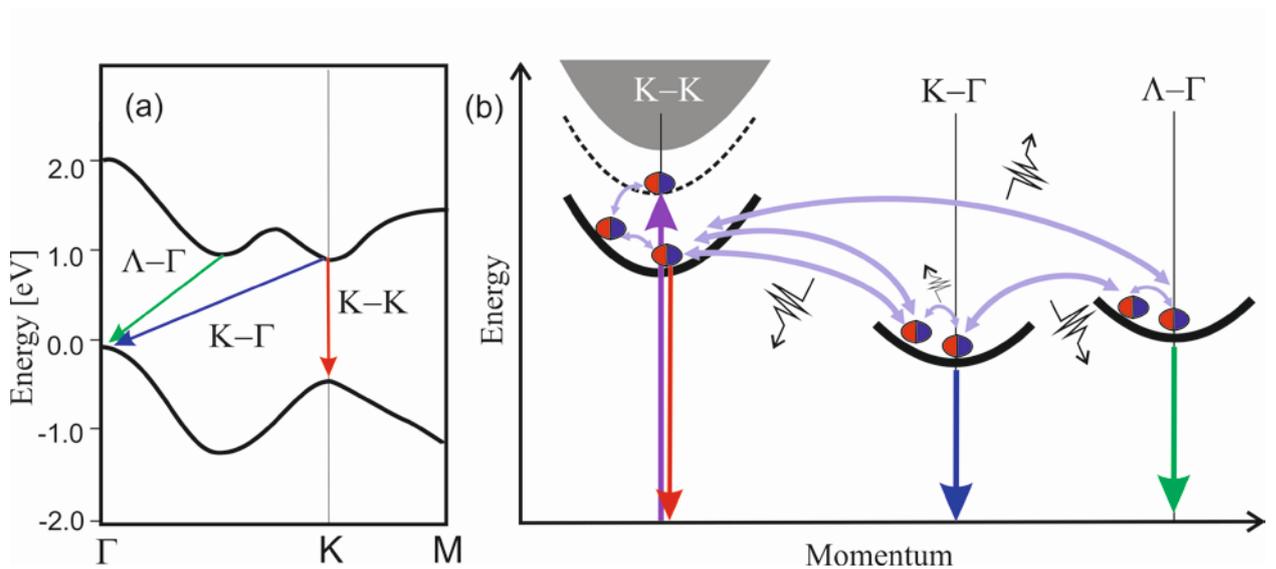

**Figure 8.** (a) Schematic band structure of multilayered 3R MoS$_2$ where only the lowest conduction and highest valence bands are depicted. (b) The scheme of formation, relaxation, and radiative recombination of the momentum direct (K–K) and indirect (K–Γ and Λ–Γ) excitonic states (red, blue, and green arrows, respectively) with quasi-resonant excitation. Absorption/emission of phonons is marked by fractured black arrows.

By globally considering these findings, we underline that the direct and indirect exciton recombination channels cannot be considered separately in the multilayered tubes and flakes. To clarify that, we present in Figure 8 the sketch of the band structure of 3R $MoS_2$ together with the scheme of exciton creation, relaxation and recombination. Obviously, when the recombination rate of strong direct excitonic resonances is high enough, the K–K transitions can recombine with emitting a photon instead of transferring the excitations towards the K–Γ or Λ–Γ states by a less probable phonon-assisted mechanism. As described in section 2, there are several factors that can promote the direct exciton emission in multi-layered systems. In particular, a quasi-resonant excitation and strain in NTs can markedly change the ratio between the direct and indirect excitonic transitions. We cannot also ignore such effects as interlayer coupling and exciton tunneling in chiral structures.[91, 92]

## 7. TMD tubes as optical resonators

The tubular microcavities supporting whispering gallery modes (WGMs) have attracted much attention due to their unique properties, such as subwavelength dimension, light polarization, and superior optical confinement due to the absence of higher-order radial modes. That makes them promising for different nanophotonics applications, including lasing.[93, 94] A variety of techniques were exploited to fabricate tubular resonators from so-called "epitaxial casting" when a core is etched in a core-shell column to direct lithography formation.[95, 96] The rolling-up strained thin films or membranes were also used to form tubular micro-resonators.[97] The high quality of such rolls provides spectra with sharp peaks of optical modes.[98] A specially designed bottlelike shape with varied numbers of layers in a wall allowed ones to study the effects of light confinement, mode splitting, and chirality.[99-101] We should underline that the studies of the microtubes and microrolles as optical resonators have mainly concerned the tubular structures made from layers of bulk materials. In the 2D family, the carbon nanotubes have solely been investigated as a part of optomechanical and hybrid cavity systems.[102- 104] To the best of our knowledge, the first demonstration of a single synthesized NT acting as a resonator has been done during our studies[28].

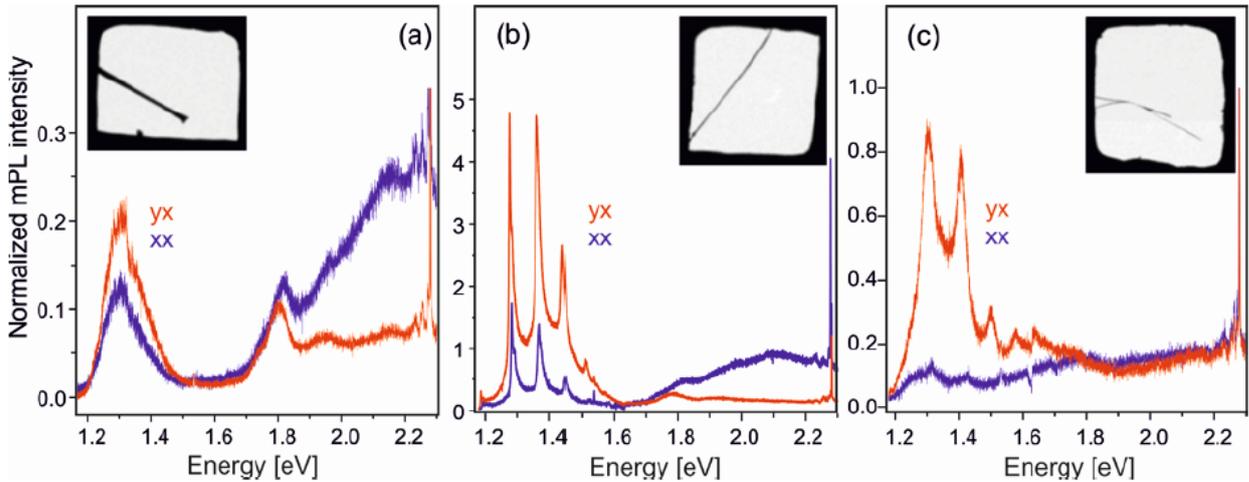

**Figure 9.** Micro-PL spectra measured in samples suspended in a mesh cup as shown in the inserts: (a) a 4-µm-thick ribbon, (b) a tube with a diameter of 2 µm, and (c) a tube with a diameter of 800 nm. The spectra normalized to the intensity of the phonon mode are recorded at room temperature in two different polarization configurations, using a 532-nm laser line with a power of 0.4 mW for excitation. The difference between higher-energy parts of the spectra is mainly due to the different light scattering along and across the tube/ribbon axes. No sharp WGM peaks are observed in the spectra of the ribbon.

The typical mPL spectrum with pronounced sharp peaks related to the optical modes is shown in Figure 6b. The energy dependence of the energy splitting between adjacent peaks, named as free spectral range (FSR), as well as its shift with a temperature rise (see Figure 6b, inset) follow the refractive index variation near the A exciton resonance, as it was observed previously for the WGMs in semiconductor resonators.[105] We found that the WGM peaks are strongly polarized along the tube axis. This effect was observed in the NTs of various diameters placed both on the silica substrate and suspended on the TEM grid (Figure 9). It is worth noting that the WGM peaks are strong in the *yx* and *xx* configurations and that their energies are similar in these two polarization configurations. They are suppressed in two other, *xy* and *yy*, configurations. That is why we present here the spectra only for the strongly different *yx* and *xy* configurations. To compare the PL intensity in different samples we normalize each PL spectrum in Figure 9 to the intensity of the phonon mode $E^1_{2g}$, not dependent on layer thickness and polarization (see Figure 4a). This normalization excludes a possible difference in excitation power densities due to accidental tube displacement and, hence, reflects the true PL efficiency.[17]

The WGMs peaks are not observed in the collapsed tubes – ribbons (Figure 9a). Thus, this is a way to distinguish the tubes from the ribbons. In the spectra of NTs with small diameters (<500 nm), the WGM related peaks are almost absent (see, e.g., the spectrum of a 400-nm NT in Figure 5a). On the contrary, the WGM peaks are very pronounced in tubes of large diameters ~2 µm. The quality factor Q=□□□□ of the tubular resonator can be as high as 100-200 (Figure 9b). As a result, the PL intensity at the WGM frequencies can be locally increased by 50 times as compared with the intensity of PL in the *xy*-polarization, which is not enhanced by such modes. Consideration of the spectra presented in Fig. 6b and Fig. 9 shows that optical modes can amplify any of the two bands associated with direct or indirect excitons.

In a tubular resonator, the spectral position of the WGMs must depend on the tube diameter and the wall thickness. The theoretical consideration of optical modes quantified in the walls of vdW NTs can be found in our recent paper.[28] Briefly, the mPL spectra of NTs were modelled for different polarization configurations using the Lorentz reciprocity theorem. After excitation by $E(\omega)_{exc}$ at the frequency ω, the exciton generation in the wall at a radius *r* from the tube center and at an angle θ is characterized by the spatial distribution $P(r\theta)$. Electric fields induced along orthogonal directions $E_\theta$ and $E_x$ determine the PL spectra in two experimental configurations $PL_{xy}(\omega)$ and $PL_{yx}(\omega)$, respectively. The exciton generation is followed by rapid energy relaxation to the recombining states. The subsequent photon emitting occurs with the conservation of real space distribution and the loss of polarization. The non-polarized PL is modulated by the strongly polarized resonator modes. We achieve the perfect modelling the energy and shape of the WGM peaks assuming one monolayer fluctuation in the wall thickness. Since the number of monolayers, *N*, in the tube wall is the fitting parameter; the modelling is used to estimate this quantity. In particular, $N \approx 45$ is derived for the microtube whose spectra are shown in Figure 6b.

The WGMs oscillate in the azimuthal direction with frequency which is dependent on the azimuthal angular momentum number, $m$, which in turn depends on the tube diameter, $D$, and $N$. The difference between $yx$-polarized and $xy$-polarized PL, which is clearly seen in all measured spectra, is explained by the different angular momentum numbers $m$ of the modes supported at certain energy in these polarizations. It is well known that the $Q$-factor increases with the rise of $m$ number.[106] Our modelling has shown that near the A exciton, the $m$ value is about 20 in the $yx$-polarization, but it is significantly less in the $xy$-polarization. In the $yx$-polarization, the PL intensities in the WGN peaks increase until the strong absorption at the resonances of direct excitons suppresses any enhancement by optical modes.

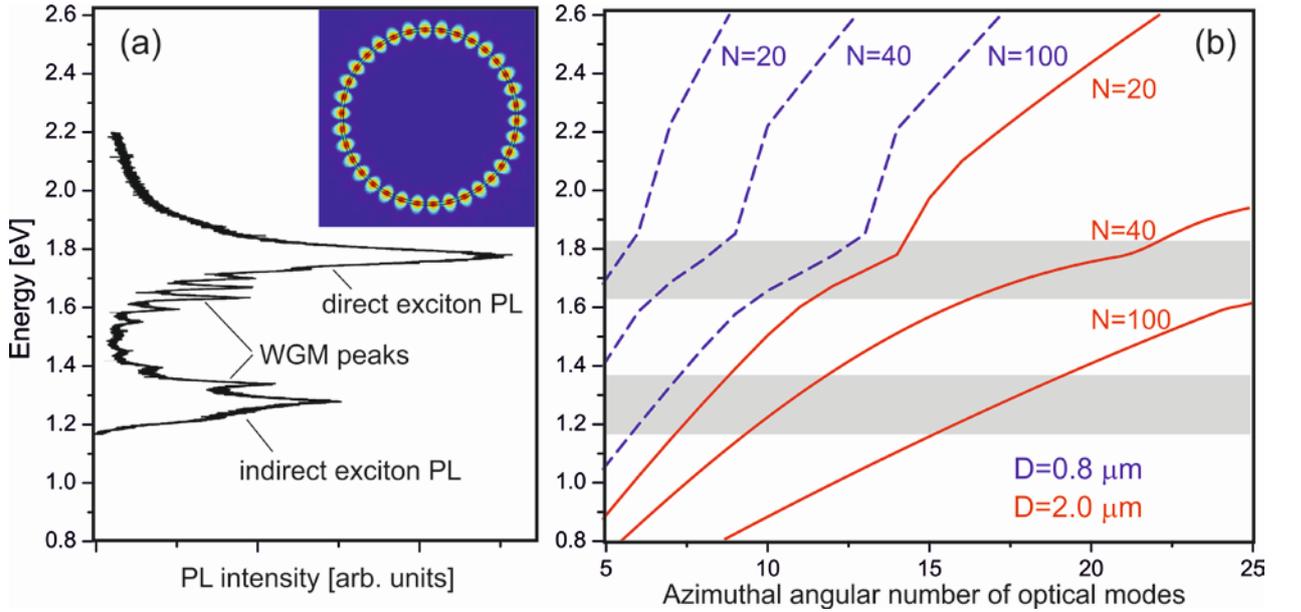

**Figure 10.** (a) Typical PL spectrum measured in $yx$-polarization in a MoS$_2$ tube with the diameter $D=2$ µm and a number of monolayers $N\approx45$. The WGM peaks are pronounced below the direct exciton resonance. The inset represents the electric field distribution in the tube wall for a WGM with azimuthal angular number $m = 17$. (b) Dependences of the optical mode energies on $m$ calculated for the tubes with $D$ of 0.8 µm (blue lines) and 2 µm (red lines) and different number of monolayers in their walls, *N*, marked in the plot. Shaded areas denote the spectral ranges of excitonic emission.

To provide a reasonable enhancement, an essential part of the mode energy should be confined inside the tube walls, which is also achieved with $m \geq 20$. To illustrate that, we show in Figure 10a the characteristic NT spectrum with the WGM peaks plotted along the vertical axis. The inset in this figure shows the confined energy distribution. Figure 10b presents the dependences of the mode energy on *m*, calculated for two diameters (0.8 and 2 µm) and different *N* (20, 40, and 100). This graph shows that the larger the diameter of the nanotube *D* is, the greater the azimuthal angular number *m* of whispering gallery modes in the spectral ranges of interest. This ensures a higher concentration of electromagnetic energy inside the tube walls. The calculated dependences also show that, for a given value of *D*, the increase in the number of layers *N* leads to the same effect because of the increasing slope of curves. Comparison of two panels (a) and (b) shows that in order to enhance the emission of direct exciton band at 1.9 eV the tubes must be of large diameter and relatively thin-walled. This is realized for the 2-µm tube with N~40-50. The optical modes in the thick-walled tubes of large diameters (almost full cylinder) can enhance only the indirect exciton PL near 1.3 eV, whereas the $MoS_2$ tubes supporting the modes with $m<15$ could provide a weak enhancement, but at higher energy, if the absorption here would not be so strong. The experimental spectra measured in micro- and nanotubes are well consistent with this simplified consideration.

## 8. Conclusions and future outlook

In summary, we present optically-active TMD micro- and nanotubes synthesized by well-developed CTR technique which ensured their excellent structural properties. Our data show that the multilayered tubes are both radiating and resonating. It is found that they exhibit emission related not only to the indirect excitons but also to direct excitons and that the direct-exciton radiative transitions can dominate the emission spectra. In the common opinion, the multilayered structures made from TMDs with indirect band gap cannot effectively radiate. Their emission is expected to be weak and hardly measurable. To elucidate the reasons of such discrepancy, we have reviewed old and recent studies of electronic band structures and optical properties of bulk TMDs and related nanostructures, including curved and tubular ones. It is found that several factors can promote the unexpected performance of excitonic emission. Among them, there are strain, interlayer distance variation, intercalated impurities, and quasi-resonant excitation. In addition, there should be some peculiar factors such as internal tubular arrangement – specific folding and chirality, as well as the fine structure of excitonic states, comprising both bright and dark excitons. Full understanding of the role of these factors requires thorough research. We believe that micro-spectroscopy with temporal resolution and with applied magnetic field, supported by theoretical considerations, can shed light on the unusual optical properties of the synthesized TMD NTs.

Concerning the resonating ability of TMD NTs, we should underline that the spectra with strong WGM peaks were measured at room temperature and that the $Q$-factor as high as several hundred is achievable in the synthesized $MoS_2$ tubes. This opens a way for their applications as effective microresonators in nanophotonics. Note that manipulations with these tubes do not require the laborious "tape" technology. These advantages and exceptional optical properties present a playground for realization of interesting nanophotonic systems where the NTs can be used for reemitting, selective enhancement, and transferring of the electromagnetic energy.

**Aknowledgements.** This work was partly supported by the Government of the Russian Federation (Project No. 14.W03.31.0011 at the Ioffe Institute). The micro-spectroscopy measurements were supported by the Russian Science Foundation (project #14-22-00107). The synthesis and structural characterization of TMD NTs were supported by Slovenian Research Agency (P1-0099). We thank A.N. Smirnov and M.V. Rakhlin for the help in micro-spectroscopy measurements; A.V. Poshakinskiy and D.R. Kazanov for WGM modelling; D.V. Rybkovskiy for preliminary DFT calculations; D.A. Kirilenko for the characterization of samples